# Transformer based Grapheme-to-Phoneme Conversion

*Sevinj Yolchuyeva[1], Géza Németh[2], Bálint Gyires-Tóth[3]*

[1,2,3]Department of Telecommunications and Media Informatics, Budapest University of Technology and Economics, Budapest, Hungary

{syolchuyeva, nemeth, toth.b}@tmit.bme.hu

## Abstract

Attention mechanism is one of the most successful techniques in deep learning based Natural Language Processing (NLP). The transformer network architecture is completely based on attention mechanisms, and it outperforms sequence-to-sequence models in neural machine translation without recurrent and convolutional layers. Grapheme-to-phoneme (G2P) conversion is a task of converting letters (grapheme sequence) to their pronunciations (phoneme sequence). It plays a significant role in text-to-speech (TTS) and automatic speech recognition (ASR) systems. In this paper, we investigate the application of transformer architecture to G2P conversion and compare its performance with recurrent and convolutional neural network based approaches. Phoneme and word error rates are evaluated on the CMUDict dataset for US English and the NetTalk dataset. The results show that transformer based G2P outperforms the convolutional-based approach in terms of word error rate and our results significantly exceeded previous recurrent approaches (without attention) regarding word and phoneme error rates on both datasets. Furthermore, the size of the proposed model is much smaller than the size of the previous approaches.

**Index Terms**: Attention mechanism; Grapheme-to-Phoneme (G2P); Transformer architecture; Multi-head attention

## 1. Introduction

Grapheme-to-phoneme conversion is an important component in TTS and ASR systems [1]. Many approaches have been proposed: the early solutions were rule-based [2], while in later works, joint sequence models for G2P conversion were introduced [3, 4]. The latter requires alignment between graphemes and phonemes, and it calculates a joint n-gram language model over sequences. The method proposed by [3] is implemented in the publicly available tool, called Sequitur.
Encoder-decoder architectures were applied in various tasks, such as neural machine translation, speech recognition, text-to-speech synthesis [1,5,6]. When combined with different attention mechanisms, it achieved state-of-the-art results in different fields. This combination was investigated by [7] for the G2P task and resulted in state-of-the-art G2P performance without explicit alignments, the phoneme error rate (PER) being 4.69% and the word error rate (WER) reaching 20.24% on CMUDict. In [1], an end-to-end TTS system (constructed entirely from deep neural networks) utilized an encoder-decoder model for the G2P task by using the multi-layer bidirectional encoder with GRU (Gated Recurrent Unit) and a deep unidirectional GRU decoder.

Convolutional neural networks have achieved superior performance compared to previous methods in large-scale image recognition [8]. Recently, encoder-decoder architectures using convolutional neural networks have been studied and applied to various Natural Language Processing (NLP) tasks [9, 10]. Convolutional neural network-based sequence-to-sequence architecture for G2P was introduced in [11]. This approach achieved a 4.81% phoneme error rate (PER) and 25.13% word error rate (WER) on CMUDict; 5.69% PER and 30.10% WER on NetTalk. The proposed model is based on convolutional layers with residual connections as an encoder and a Bi-LSTM decoder.
In sequence-to-sequence learning, the decoding stage is usually carried out sequentially, one step at a time from left to right and the outputs from the previous steps are used as decoder inputs [12]. Sequential decoding can negatively influence the results, depending on the task and the model. The non-sequential greedy decoding (NSGD) method for G2P was studied in [12], and it was also combined with a fully convolutional encoder-decoder architecture. That model achieved 5.58% phoneme and 24.10% word error rates on CMUDict, which included multiple pronunciations and without stress labels.
Multilingual G2P models are used for multilingual speech synthesis [26]. In [13] monolingual G2P (MoG2P) and multilingual G2P (MuG2P) conversions were proposed, and experiments were conducted in four languages (Japanese, Korean, Thai, and Chinese) with both language-dependent and -independent trainings. Moreover, a neural sequence-to-sequence approach to G2P was presented, which is trained on spelling–pronunciation pairs in hundreds of languages [14, 23]. The proposed system shared a single encoder and decoder across all languages, allowing it to utilize the intrinsic similarities between different writing systems.
Transformer networks are based on an encoder-decoder architecture and account the representations of their input and output without using recurrent or convolutional neural networks (CNN) [15, 16]. First, transformer networks were used for neural machine translation, and they achieved state-of-the-art performance on various datasets. In [15], it was shown that transformers could be trained significantly faster than recurrent or convolutional architectures for machine translation tasks.
According to our knowledge, our approach is the first study that applies the transformer for G2P conversion. In this paper, we present transformers with different structures and analyse their advantages and disadvantages for G2P task. Our main goal was to achieve and surpass (if possible) the accuracy of previous models and to reduce the required resources of training.
The rest of the paper is organized as follows: Section 2 describes the transformer architecture for the G2P conversion task. Datasets, training processes, the evaluation of the proposed models are presented in Section 3. Evaluation and results are described in Section 4, and finally, conclusions are drawn in Section 5.

## 2. Proposed architecture

Encoder-decoder based sequence to sequence learning (seq2seq) has made remarkable progress in recent years. The main idea of these approaches has two main stages: first, the encoder converts the input sequence to a vector; second, the output sequence is generated based on the learned vector representation by using the decoder. For both encoder and decoder, different network architectures have been investigated [5, 18].

The transformer is organized by stacked self-attention and fully connected layers for both the encoder and the decoder [15], as shown in the left and right halves of Figure 1, respectively. Self-attention, sometimes called intra-attention, is an attention mechanism relating different positions of a single sequence to compute its internal representation.

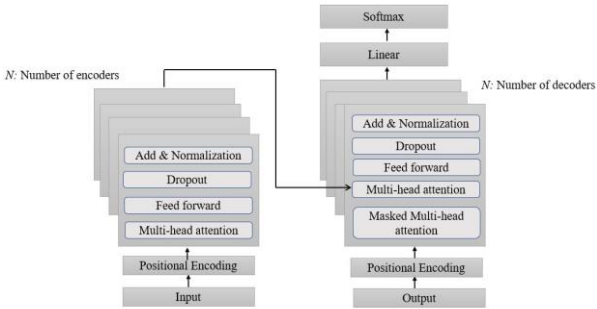

Figure 1: *The framework of the proposed model.*

Without using any recurrent layer, positional encoding is added to the input and output embeddings [16]. The positional information provides the transformer network with the order of input and output sequences.

The encoder is composed of a stack of $N$ identical blocks, and each block has two layers. The first is the multi-head attention layer, which is several attention layers used in parallel. The second is a fully connected position-wise feed forward layer. These layers are followed by dropout and normalization layers [24]. The decoder is composed of a stack of $N$ identical blocks, and each block has three layers. The first layer is the multi-head attention mechanism with masked [17]. This mechanism helps the model to generate the current phoneme using only the previous phonemes. The second layer is a multi-head attention layer without the masked. It performs the multi-head attention over the output of the first layer. The third layer is fully connected. These layers are followed by normalization [24] and dropout layers [25]. At the top, there is the final fully connected layer with linear activation which is followed by softmax output.

An attention function was described as mapping a query and a set of key-value pairs to an output, where the query *(Q)*, keys *(K)*, values *(V)*, and output are all vectors [15]. A multi-head attention mechanism builds upon scaled dot-product attention, which computes on a query $Q$, key $K$ and a value $V$ (the dimension of queries and keys is $d_k$ and values of dimension is $d_v$):

$$Attention(Q,K,V) = softmax\left(\frac{QK^T}{\sqrt{d_k}}\right)V \quad (1)$$

where the scalar $1/\sqrt{d_k}$ is used to prevent softmax function into regions that have very small gradients.

Instead of performing a single attention function multi-head attention obtains $h$ (parallel attention layers or heads) for learning different representations, compute scaled dot-product attention for each representation, concatenate the results, and project the concatenation with a feedforward layer. Finally, $d_m$ dimensional outputs are obtained. The multi-head attention is shown as follows [15]:

$$h_i = Attention(QW_i^Q, KW_i^K, VW_i^V) \quad (2)$$

$$MultiHead(Q,K,V) = Concat(h_1, h_2, .., h_h)W^O \quad (3)$$

where the projections are parameter matrices $W_i^Q \in R^{d_m \times d_k}, W_i^K \in R^{d_m \times d_k}, W_i^V \in R^{d_m \times d_v}$ and $W^O \in R^{hd_v \times d_m}$. Each head in multi-head attention learns individual sequence dependency, and this allows the model to attend to information from different representation subspaces. So, it increases the power of the attention with no computational overhead.

## 3. Experiments

In this section, we introduce the datasets and then describe the implementation details.

### 3.1. Datasets

For evaluation, the CMU pronunciation[1] and NetTalk datasets were used. These datasets have been frequently chosen by researchers [7, 18]. The train and test split was the same as found in [7, 18, 22], thus, the results are directly comparable. CMUDict contains a 106,837-word training set and a 12,000-word test set (reference data). 2,670 words are used as development (validation) set. There are 27 graphemes (uppercase alphabet symbols plus the apostrophe) and 41 phonemes in this dataset. NetTalk contains 14,851 words for training, 4,951 words for testing and does not have a predefined validation set. There are 26 graphemes (lowercase alphabet symbols) and 52 phonemes in this dataset.

We use **<START>, <END>** tokens as beginning-of-graphemes (beginning-of-phonemes), end-of-graphemes (end-of-phonemes) tokens and **<PAD>** token in both datasets.

### 3.2. Software and hardware details

NVidia Titan Xp (12 GB) and NVidia Titan X (12 GB) GPU cards hosted in two i7 desktop servers with 32GB RAM served for training and inference. For training and evaluation, the Keras[2] deep learning framework with TensorFlow[3] backend was our environment.

---

[1] http://www.speech.cs.cmu.edu/cgi-bin/cmudict
[2] https://keras.io/
[3] https://www.tensorflow.org/

### 3.3. Training

By adding <**START**>, <**END**> and <**PAD**> tokens to the input and output, the length of the longest input and output was fixed to 24. We completed shorter input and output sequences with the <**PAD**> token to make their length equal in both training and development sets. For the test set, padding was not applied.

We applied two embeddings which represent the encoder (grapheme) and decoder (phoneme) sides, respectively. The encoder and decoder embeddings had a great influence on the results. The size of the embeddings is 128, and the dimension of the inner-layer is 512. We used Adam as optimizer [19]. The initial learning rate was set to 0.0002. If the performance (PER for G2P conversion) on the validation set has not improved for 50 epochs, the learning rate was multiplied by 0.2. We apply layer normalization and dropout in all models. The dropout rate of encoder and decoder is set to 0.1. Batch size is 128 for CMUDict, 64 for NetTalk. We have investigated three transformer architectures, with 3 encoder and decoder layers (it is called Transformer 3x3 in Table 2), 4 encoder and decoder layers (it is called Transformer 4x4 in Table 2) and 5 encoder and decoder layers (it is called Transformer 5x5 in Table 2).

We employed $h = 4$ parallel attention layers in all proposed models, and $Q$, $K$ and $V$ have the same dimension of $d_m$, so that $d_v = d_k = d_m = 128$ and $d_m/h = 32$. Due to the reduced dimension of each head, the total computational cost is similar to that of single-head attention with full dimensionality.

Other parameters used in training are defined in Table 1.

Table 1: *Training parameters.*

| Parameters | Number |
|---|---|
| Encoder layers ($N$) | 3/4/5 |
| Decoder layers ($N$) | 3/4/5 |
| Params in one encoder | 256 |
| Params in one decoder | 256 |
| Dropout | 0.1 |
| Batch size | 128/64 |
| Adam optimizer | $\beta_1 = 0.9$, $\beta_2 = 0.998$ |

### 3.4. Inference

During inference, the phoneme sequence (written pronunciation form of given grapheme sequence) will be generated one-by-one at a time.

The sequence begins with the start token <**START** >, and we generate the first phoneme by the highest probability. Then, this phoneme is fed back into the network to generate the next phoneme. This process is continued until the end token <**END**> is reached, or the maximal length terminates the procedure. Beam search was not applied in this work.

## 4. Evaluation and results

We use the following common evaluation metrics for G2P:

*Phoneme Error Rate (PER)* is the Levenshtein distance between the predicted phoneme sequences and the reference phoneme sequences, divided by the number of phonemes in the reference pronunciation [20]. In case of multiple pronunciation samples for a word in the reference data, the sample that has the smallest distance to the candidate is used.

*Word Error Rate (WER)* is the percentage of words in which the predicted phoneme sequence does not exactly match any reference pronunciation, the number of word errors is divided by the total number of unique words in the reference.

Table 2: *Results on the CMUDict and NetTalk dataset.*

| Dataset | Model | PER | WER | Time [s] | Model size |
|---|---|---|---|---|---|
| CMUDict | Transformer 3x3 | 6.56 | 23.9 | 76 | 1.49M |
| | Transformer 4x4 | **5.23** | **22.1** | 98 | 1.95M |
| | Transformer 5x5 | 5.97 | 24.6 | 126 | 2.4M |
| NetTalk | Transformer 3x3 | 7.01 | 30.67 | 33 | 1.50M |
| | Transformer 4x4 | **6.87** | **29.82** | 39 | 1.96M |
| | Transformer 5x5 | 7.72 | 31.16 | 48 | 2.4M |

Table 3: *Results on the CMUDict and NetTalk datasets.*

| Data | Method | PER (%) | WER (%) | Model size |
|---|---|---|---|---|
| NetTalk | Joint sequence model [3] | 8.26 | 33.67 | N/A |
| | Encoder-decoder with global attention [7] | 7.14 | 29.20 | N/A |
| | Encoder CNN with res. conn, decoder Bi-LSTM (Model 5) [11] | 5.69 | 30.10 | 14.5 M |
| | **Transformer 4x4** | **6.87** | **29.82** | **1.95 M** |
| CMUDict | Encoder-decoder LSTM [18] | 7.63 | 28.61 | N/A |
| | Joint sequence model [3] | 5.88 | 24.53 | N/A |
| | Combination of sequitur G2P and seq2seq-attention and multitask learning [21] | 5.76 | 24.88 | N/A |
| | Deep Bi-LSTM with many-to-many alignment [23] | 5.37 | 23.23 | N/A |
| | Joint maximum entropy (ME) n-gram model [4] | 5.9 | 24.7 | N/A |
| | Encoder CNN, decoder Bi-LSTM (Model 5) [11] | 4.81 | 25.13 | 14.5M |
| | End-to-end CNN (Model 4) [11] | 5.84 | 29.74 | 7.62M |
| | Encoder-decoder LSTM with attention (Model 1) [11] | 5.68 | 28.44 | 12.7M |
| | **Transformer 4x4** | **5.23** | **22.1** | **2.4M** |

Table 4. *Examples of errors predicted by Transformer 4x4 and [11].*

|  | Example 1 | Example 2 | Example 3 |
|---|---|---|---|
| Original word | NATIONALIZATION | KORZENIEWSKI | GRANDFATHERS |
| Reference | N AE SH AH N AH L AH Z EY SH AH N | K AO R Z AH N UW F S K IY | G R AE N D F AA DH ER Z |
| Prediction of CNN based model [11] (Model 5) | N AE SH AH N AH L AH <u>EY EY</u> SH AH N | K AO R Z <u>N N N</u> UW <u>S</u> K IY | G R AE N D <u>AA DH DH ER</u> |
| Transformer 4x4 | N AE SH <u>N AH</u> L AH Z EY SH AH N | K <u>ER</u> Z AH N <u>UW S</u> K IY | G R AE <u>N F</u> AA DH ER Z |

After training the model, predictions were run on the test dataset. The results of the evaluation on CMUDict and NetTalk are shown in Table 2. The first and second columns show the dataset and the applied architecture, respectively. The third and fourth columns show the PER and WER values. The fifth column of Table 2 contains the average sum of training and validation time of one epoch. The last column presents information about the number of parameters (weights). According to the results, Transformer 4x4 (4 layers encoder and 4 layers decoder) outperforms Transformer 3x3 (3 layers encoder and 3 layers decoder). Contrary to expectations Transformer 5x5 (5 layers encoder and 5 layers decoder) didn't outperform Transformer 4x4 (4 layers encoder and 4 layers decoder). Increasing the numbers of encoder-decoder layers leads to much more training parameters. In the G2P task, similar complexity to NMT (neural machine translation) can be rarely permitted. The high number of parameters sometimes does not even result in better performance. In term of PER, Transformer 5x5 is better than Transformer 3x3 on CMUDict but didn't exceed Transformer 4x4, Transformer 3x3 in the point of WER on both CMUDict and NetTalk.

During the experiments, we did not observe significant performance improvements when the number of encoder-decoder was increased.

In Table 3, the performance of the Transformer 4x4 model with previously state-of-the-art results is compared on both CMUDict and NetTalk databases. The first column shows the dataset, the second column presents the method used in previous solutions with references, PER and WER columns tell the results of the referred models, and the last column presents information about the number of parameters (weights). According to Table 3, our proposed model reached competitive results for both PER and WER. For NetTalk, we are able to exceed previous results significantly. We should point out that the results of the Transformer 4x4 model are close to encoder CNN with residual connections, decoder Bi-LSTM model obtained by [11] regarding PER, but WER is better in the proposed model. Moreover, the number of parameters of the convolutional layers with residual connections as encoder and Bi-LSTM as the decoder is 14.5M, encoder-decoder LSTM and encoder-decoder Bi-LSTM have 12.7M and 33.8M, respectively [11]. Both the Transformer 4x4 and the Transformer 3x3 have fewer parameters than the previously mentioned models.

When comparing Transformer 4x4 and encoder CNN, decoder Bi-LSTM model [11], there is an interesting contravention between PER and WER. Although PER is smaller, WER is higher in the encoder CNN, decoder Bi-LSTM model than Transformer 4x4. As mentioned in [11], there were twice as many words with only one phoneme error than words which have two phoneme errors in the result of encoder CNN decoder Bi-LSTM model, and it affected the growth of WER. In contrast, in Transformer 4x4 the number of words with only one phoneme error is not too much. Regarding the types of error when generating phoneme sequences, in the CNN encoder, Bi-LSTM decoder, some phonemes are unnecessarily generated multiple times. For example, for the word KORZENIEWSKI, reference is [ K AO R Z AH N UW F S K IY], the prediction of CNN encoder, Bi-LSTM decoder for this word is [K AO R Z N N N UW S K IY], where the character N was generated three times. But the prediction of Transformer 4x4 for this word is [K ER Z AH N UW S K IY], where 1 failed phoneme (ER) and 2 forgotten phonemes (R, F) appear. Example 1 and Example 3 in Table 4 also show the type of errors for Transformer 4x4 and CNN encoder Bi-LSTM decoder.

## 5. Conclusions

We investigated a novel transformer architecture for the G2P task. Transformer 3x3 (3 layers encoder and 3 layers decoder), Transformer 4x4 (4 layers encoder and 4 layers decoder), and Transformer 5x5 (5 layers encoder and 5 layers decoder) architectures were presented including experiments on CMUDict and NetTalk. We evaluated PER and WER, and the results of the proposed models are very competitive with previous state-art results. The number of parameters (weights) of all proposed models is less than the CNN and the recurrent models. As a result, the time consumption of training process decreased.

In future research, we intend to study the application of the proposed method in the field of end-to-end TTS synthesis.

## 6. Acknowledgements

The research presented in this paper has been supported by the European Union, co-financed by the European Social Fund (EFOP-3.6.2-16-2017-00013, Thematic Fundamental Research Collaborations Grounding Innovation in Informatics and Infocommunications), by the BME-Artificial Intelligence FIKP grant of Ministry of Human Resources (BME FIKP-MI/SC), by Doctoral Research Scholarship of Ministry of Human Resources (ÚNKP-18-4-BME-394) in the scope of New National Excellence Program, by János Bolyai Research Scholarship of the Hungarian Academy of Sciences, by the AI4EU project (No 825619), and the DANSPLAT project (Eureka 9944). We gratefully acknowledge the support of NVIDIA Corporation with the donation of the Titan Xp GPU used for this research.
We are grateful to Stan Chen for providing the dataset of NetTalk.